\documentclass{article}
\usepackage{emulateapj,epsfig}

\slugcomment{ApJ, in press}

\makeatletter

\newenvironment{inlinefigure}{%
\def\@captype{figure}%
\noindent\begin{minipage}{0.999\linewidth}\begin{center}}
{\end{center}\end{minipage}\smallskip}
\makeatother

\def\ltsima{$\; \buildrel < \over \sim \;$}
\def\simlt{\lower.5ex\hbox{\ltsima}}
\def\gtsima{$\; \buildrel > \over \sim \;$}
\def\simgt{\lower.5ex\hbox{\gtsima}}

\begin{document}

\title{High-Redshift Gamma-Ray Bursts from Population~III Progenitors}

\author{Volker Bromm \altaffilmark{1} and Abraham Loeb\altaffilmark{2}}

\altaffiltext{1}{Department of Astronomy, University of Texas, Austin, TX
78712; vbromm@astro.as.utexas.edu}

\altaffiltext{2}{Department of Astronomy, Harvard University, 
Cambridge, MA 02138; aloeb@cfa.harvard.edu}

\begin{abstract}
Detection of gamma-ray bursts (GRBs) from redshifts $z\ga 7$ would open a
new window into the earliest epoch of cosmic star formation. We construct
separate star formation histories at high redshifts for normal (Pop~I and
II) stars, and for predominantly massive (Pop~III) stars. Based on these
separate histories, we predict the GRB redshift distribution to be observed
by the {\it Swift} mission.  Regardless of whether Pop~III progenitors are
able to trigger GRBs, we find that a fraction $\sim 10$\% of all bursts
detected by {\it Swift} will originate at $z\ga 5$. This baseline
contribution is due to Pop~I/II star formation which must have extended out
to high redshifts in rare massive galaxies that were enriched by heavy
elements earlier than the typical galaxies. In addition, we consider the
possible contribution of Pop~III progenitors to the observable GRB rate.
Pop~III stars are viable progenitors for long-duration GRBs which are
triggered by the collapsar mechanism, as long as they can lose their outer
envelope through mass transfer to a companion star in a close binary.  We
find that the likelihood of Pop~III binaries to satisfy the conditions
required by the collapsar mechanism could be enhanced significantly
relative to Pop~I/II binaries. If Pop~III binaries are common, {\it Swift}
will be the first observatory to probe Pop~III star formation at redshifts
$z\ga 7$.
\end{abstract}


\section{Introduction}

The first stars in the universe, so-called Population~III (hereafter
Pop~III), formed out of metal-free gas at the end of the cosmic dark ages
or redshifts $z\ga 10$ (for reviews, see, e.g., Barkana \& Loeb 2001; Bromm
\& Larson 2004; Ciardi \& Ferrara 2005; Glover 2005).  These stars are
predicted to have been predominantly very massive with $M_{\ast}\ga 100
M_{\odot}$ (Bromm, Coppi, \& Larson 1999, 2002; Abel, Bryan, \& Norman
2002; Nakamura \& Umemura 2001), and to have left a mark on the thermal and
chemical evolution of the intergalactic medium (IGM). First, their
predicted high surface temperatures (e.g., Bond, Arnett, \& Carr 1984)
implies that they may have been efficient sources of ionizing photons
(e.g., Tumlinson \& Shull 2000; Bromm, Kudritzki, \& Loeb 2001b; Schaerer
2002). A contribution from Pop~III stars to the reionization of the IGM may
be required to account for the large optical depth to Thomson scattering
inferred by the {\it Wilkinson Microwave Anisotropy Probe} (WMAP; Kogut et
al. 2003) during its first year of operation (e.g., Cen 2003; Wyithe \&
Loeb 2003a, 2003b).  Second, since the stellar evolutionary timescale for
massive Pop~III stars is short $\sim 10^{6}$~yr, the resulting initial
enrichment of the IGM with heavy elements could have occurred rather
promptly (e.g., Mori, Ferrara, \& Madau 2002; Bromm, Yoshida, \& Hernquist
2003; Wada \& Venkatesan 2003; Yoshida, Bromm, \& Hernquist 2004).

Gamma-Ray Bursts (GRBs) offer unique prospects for probing the cosmic star
formation (Totani 1997; Wijers et al. 1998; Blain \& Natarajan 2000;
Porciani \& Madau 2001; Bromm \& Loeb 2002; Hernquist \& Springel 2003;
Mesinger, Perna, \& Haiman 2005;
Natarajan et al. 2005) as well as the IGM (Loeb 2003; Barkana \& Loeb 2004;
Gou et al. 2004; Inoue, Omukai, \& Ciardi 2005; Ioka \& M\'{e}sz\'{a}ros
2005) at redshifts $z\ga 7$, beyond the current horizon of galaxy and
quasar surveys. GRBs are the brightest electromagnetic explosions in the
universe (for recent reviews, see, van Paradijs, Kouveliotou, \& Wijers
2000; M\'{e}sz\'{a}ros 2002; Piran 2004) and their emission is detectable
out to $z\ga 10$. The detectability of their gamma-ray emission (Lamb \&
Reichart 2000) allows simultaneous monitoring of a major portion of the
sky, while the detectability of their afterglow (Ciardi \& Loeb 2000)
allows to determine their high redshift from the appearance of the
intergalactic Ly$\alpha$ absorption trough in the infrared (Loeb 2003;
Barkana \& Loeb 2004).  The popular collapsar model for the central engine
of long-duration GRBs (MacFadyen, Woosley, \& Heger 2001 and references
therein), involves the collapse of a massive star to a black hole
(BH). This model naturally explains the observed association of
long-duration GRBs with star-forming regions (e.g., Fruchter et al. 1999;
Djorgovski et al. 2001; Bloom, Kulkarni, \& Djorgovski 2002a) and the Type
Ib/c supernova signature on the spectra of rapidly decaying afterglow
lightcurves (e.g., Bloom et al. 2002b; Hjorth et al. 2003; Matheson et
al. 2003; Stanek et al. 2003).

Because of their high characteristic masses, Pop~III stars could
potentially lead to high redshift GRBs. The recently launched {\it Swift}
satellite\footnote{See http://swift.gsfc.nasa.gov} (Gehrels et al. 2004) is
ideally suited to utilize this novel window into the high-redshift
universe.  In the following sections we address the underlying question:
{\it which fraction of high-redshift bursts could originate from Pop~III
progenitors?}  The actual fraction and distribution of high-$z$ GRBs to be
measured by {\it Swift}, might reflect the absence or presence of the
potential Pop~III contribution.

The organization of the paper is as follows. In \S~2 we describe our cosmic
star formation model, in particular determining the Pop~III mode at the
highest redshifts. The resulting GRB redshift distribution is calculated
in \S~3, together with a discussion of plausible GRB progenitors.
Finally, we address the implications of our results in \S~4.

\section{Star Formation Modes at High Redshifts}

We note that GRBs are expected to exist at redshifts $z\ga 7$ even in the
absence of any true Pop~III contribution.  This is due to the rapid
enrichment of the IGM with heavy elements dispersed by the first supernovae
(SNe) beginning at $z\ga 20$ (e.g., Loeb \& Haiman 1997; Madau, Ferrara, \&
Rees 2001; Bromm et al. 2003; Furlanetto \& Loeb 2003; Scannapieco,
Schneider, \& Ferrara 2003).  Once a given region of the universe has been
enriched beyond a critical metallicity, $Z_{\rm crit}\sim
10^{-3.5}Z_{\odot}$, the mode of star formation is predicted to shift from
high-mass Pop~III stars to the lower-mass Pop~I and II cases (e.g., Omukai
2000; Bromm et al. 2001a; Schneider et al. 2002; Bromm \& Loeb 2003b;
Mackey, Bromm, \& Hernquist 2003; Schneider et al. 2003).  The mass
fraction of super-critical gas is rapidly growing toward lower redshift,
and GRBs could be formed in the conventional way from metal-enriched Pop~I
and II progenitors at high-$z$. In the following discussion, we first
construct the total cosmic star formation rate (SFR), and subsequently
decompose the total SFR into separate Pop~I/II and Pop~III components.

\subsection{Star Formation History}

Our model for the total cosmic SFR closely follows that of Bromm \& Loeb
(2002), and we here only briefly describe the key assumptions.  The
abundance and merger history of the cold dark matter (CDM) halos is
described by the extended Press-Schechter formalism (Lacey \& Cole
1993). We assume that the IGM has a two-phase structure, consisting of
neutral and ionized hydrogen phases. The reionization of the IGM was likely
an extended process, occurring over $6\la z \la 20$ (e.g., Cen 2003;
Wyithe \& Loeb 2003a; Sokasian et al. 2004;
Furlanetto \& Loeb 2005). To bracket the possibilities, we consider
two reionization redshifts, $z_{\rm reion}\approx 7$ and $17$, where
$z_{\rm reion}$ corresponds to an ionization filling fraction by volume of
$\sim50$\%.  In each case, reionization is spread out over a range in
redshifts, $\Delta z/(1+z)\simeq 1$.

Within each phase of the IGM, stars are able to form in two different ways.
The first mechanism pertains to primordial, metal-free, gas. Such gas
undergoes star formation provided that it accretes onto a dark matter halo
with a sufficiently deep gravitational potential well or equivalently a
mass above a minimum value. For the neutral medium, this minimum mass is
set by the requirement that the gas will be able to cool.  Radiative
cooling by molecular hydrogen (H$_2$) allows star formation in halos with a
virial temperature $T_{\rm vir}\ga 500$ K, while atomic cooling dominates
for halos with $T_{\rm vir}\ga 10^{3.9}$ K.  Since H$_2$ can be easily
photo-dissociated by photons below the Lyman-limit, its significance in the
cosmic star formation history is unclear (e.g. Bromm \& Larson 2004 and
references therein), and so we only show results without H$_2$ cooling
in this paper. We note, however, that in the limiting case of
negligible H$_2$ photodissociation feedback, the cosmic star formation
rate at $z\ga 15$ could be larger by one order of magnitude than the
purely atomic cooling case discussed here.

\begin{inlinefigure}
\resizebox{9cm}{!}{\includegraphics{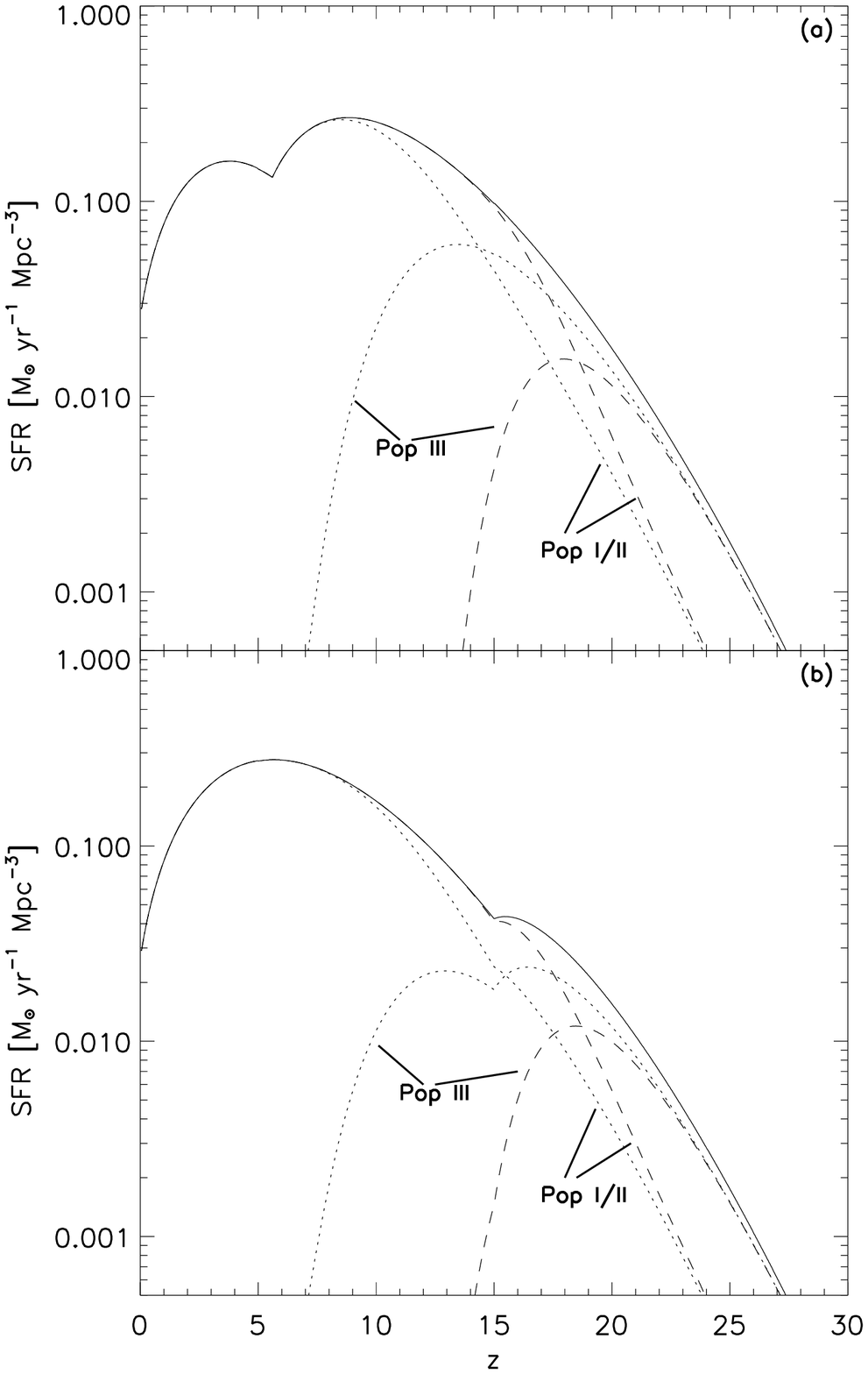}}
\caption{Cosmic comoving star formation rate (SFR) in units of
$M_{\odot}$~yr$^{-1}$~Mpc$^{-3}$, as a function of redshift.  We assume
that cooling in primordial gas is due to atomic hydrogen only, and the star
formation efficiency is $\eta_\ast=10\%$.  {\it (a)} Late reionization
($z_{\rm reion}\approx 7$).  {\it Solid line:} Total comoving SFR.  {\it
Dotted lines:} Contribution to the total SFR from Pop~I/II and Pop~III for
the case of weak chemical feedback.  {\it Dashed lines:} Contribution to
the total SFR from Pop~I/II and Pop~III for the case of strong chemical
feedback.  {\it (b)} Early reionization ($z_{\rm reion}\approx 17$).  We
adopt the same convention for the lines as in panel {\it (a)}.  In all
cases, Pop~III star formation is restricted to high redshifts, but extends
over a significant range, $\Delta z\sim 10-15$.
\label{fig1}
}
\end{inlinefigure}

For the ionized medium, on the other hand, the minimum threshold mass is
given by the Jeans mass, 
since the infall of gas and the subsequent formation of stars requires that
the gravitational force of the dark matter halo be greater than the
opposing pressure force on the gas. After reionization, the IGM is
photo-heated to temperatures $\ga 10^{4}$K, leading to a dramatic increase
in the Jeans mass. We model the suppression of gas infall according to
results from spherically-symmetric collapse simulations (see Bromm \& Loeb
2002 for details).
In calculating the late reionization case ($z_{\rm reion}\approx 7$), we employ
the prediction by Thoul \& Weinberg (1996) that gas infall is completely
suppressed in halos with circular velocities $v_c \la$ 35 km s$^{-1}$.
For the early reionization case ($z_{\rm reion}\approx 17$), however, we use
the recent work by Dijkstra et al. (2004), showing that 
the infall suppression due to photo-ionization heating could be
much less severe in the high-redshift universe. Specifically, we
assume that in this latter case infall is completely suppressed only
in halos with circular velocities $v_c \la$ 10 km s$^{-1}$.

Within our model, the second mechanism to form stars occurs in gas that has
experienced a previous burst of star formation, and is therefore already
somewhat enriched with heavy elements.  Such gas, residing in a halo of
mass $M_{1}$, can undergo induced star formation triggered by a merger with
a sufficiently massive companion halo of mass $M_{2}>0.5 M_{1}$.  We
finally assume that stars form with an efficiency of $\eta_\ast\sim 10\%$,
independent of redshift and regardless of whether the gas is primordial or
pre-enriched. This efficiency yields roughly the correct fraction of
$\Omega_{\rm B}$ found in stars in the present-day universe.  Figure 1
shows the resulting total star formation histories.  It is evident that
there are two distinct epochs of cosmic star formation, one at $z\sim 3$,
and a second one at $z\sim 8$ for late reionization, whereas there is only
a single, extended peak at
$z\sim 5$ for early reionization.

\subsection{Population III Star Formation}

To determine the fraction of the total SFR contributed by Pop~III stars, we
have to identify those halos that cross the atomic cooling threshold for
the first time. In addition, we require that the collapse takes place in a
region of the IGM which is not yet enriched with heavy elements from
previous episodes of star formation. Here, we adopt the formalism developed
in Furlanetto \& Loeb (2005) who derived the redshift-dependent probability
that a newly collapsed halo forms out of pristine gas (see their Fig.~2).
This probability crucially depends on the efficiency with which the newly
created metals are dispersed into the IGM via SN driven winds.  To bracket
the range of possibilities, we consider the cases of weak and strong
chemical feedback, corresponding to winds experiencing large and small
radiative losses, respectively\footnote{The weak and strong feedback cases
correspond to $K_{w}^{1/3}=1/3$~and~1 in equation (4) of Furlanetto \& Loeb
(2005).}.

As can be seen in Figure~1, Pop~III star formation is limited to the
highest redshifts, but in each case extends over a substantial range in
redshift: $\Delta z\sim 10$ for strong chemical feedback, and
$\Delta z\sim 15$ for weak feedback. The Pop~III histories are
rather similar for both early and late reionization. The suppression
of gas infall for the early reionization case (with $v_c \la$ 10 km s$^{-1}$),
would have a much more pronounced effect on halos that cool via H$_2$, because
of their shallower potential wells.

\section{GRB Redshift Distribution}

Next we will predict the GRB redshift distribution for flux-limited
surveys, distinguishing between the contributions from Pop~I/II and Pop~III
star formation. In particular, we will focus on the existing {\it Swift}
satellite, which is capable of making the most detailed determination of
the GRB redshift distribution to date.

\subsection{Population I/II Contribution}

Assuming that the formation of GRBs follows closely the cosmic star
formation history with no
cosmologically--significant time delay (e.g, Conselice et al. 2005),
we write
for the number of all GRB events per comoving volume per time, regardless
of whether they are observed or not: $\psi_{\rm GRB}^{\rm true}
(z)=\eta_{\rm GRB}\times \psi_{\ast}(z)$, where $\psi_{\ast}(z)$ is the
cosmic SFR, as calculated in \S 2. The efficiency factor, $\eta_{\rm GRB}$,
links the formation of stars to that of GRBs, and is in principle a
function of redshift as well as the properties of the underlying stellar
population.  The stellar initial mass function (IMF) is predicted to differ
fundamentally for Pop~I/II and Pop~III (e.g., Bromm \& Larson 2004 and
references therein).  The GRB efficiency factor will depend on the fraction
of stars able to form BHs, and consequently on the IMF (see \S 3.2). Here,
we assume that $\eta_{\rm GRB}$ is constant with redshift for Pop~I/II star
formation, whereas Pop~III stars may be characterized by a different
efficiency.

The number of bursts detected by any given instrument depends on the
instrument-specific flux sensitivity threshold and on the poorly-determined
isotropic-luminosity function (LF) of GRBs (see, e.g., Schmidt 2001; Sethi
\& Bhargavi 2001; Norris 2002).  In order to ascertain what {\it Swift} is
expected to find, we modify the true GRB event rate as follows:
\begin{equation}
\psi_{\rm GRB}^{\rm obs}(z)=\eta_{\rm GRB} \psi_{\ast}(z)
\int_{L_{\rm lim}(z)}^{\infty}p(L) {\rm d}L \mbox{\ \ .}
\end{equation}
Here, $p(L)$ is the GRB LF with $L$ being the intrinsic,
isotropic-equivalent
photon luminosity (in units of photons s$^{-1}$). If
$f_{\rm lim}$ denotes the sensitivity threshold of a given instrument (in
photons s$^{-1}$ cm$^{-2}$), then the minimum luminosity is:
\begin{equation}
L_{\rm
lim}(z)= 4 \pi d_{L}^{2} f_{\rm lim}(1+z)^{\alpha -2}\mbox{\ ,}
\end{equation}
where $d_{L}$ is the luminosity
distance to a source at redshift $z$, and $\alpha$ the intrinsic high-energy
spectral index (Band et al. 1993). For definiteness, we assume $\alpha = 2$,
which gives a reasonable fit to the observed burst spectra (see Band et al. 1993
for a detailed discussion).
We here use the same lognormal LF and the same parameters as described in
Bromm \& Loeb (2002).

In Bromm \& Loeb (2002), we predicted that $\sim 25$\% of all bursts
observed by {\it Swift} would originate at $z\ge 5$. This estimate was
based on a flux threshold of $f_{\rm lim}=0.04$ photons~s$^{-1}$~cm$^{-2}$.
Based on the first few months of actual observations by {\it Swift}, the
sensitivity limit has recently been revised upward to $f_{\rm lim}=0.2$
photons s$^{-1}$~cm$^{-2}$, comparable to the older BATSE experiment (e.g.,
Berger et al. 2005). Using this revised flux limit, we predict that $\sim
10$\% of all {\it Swift} GRBs will originate at $z\ge 5$.

\begin{inlinefigure}
\resizebox{9cm}{!}{\includegraphics{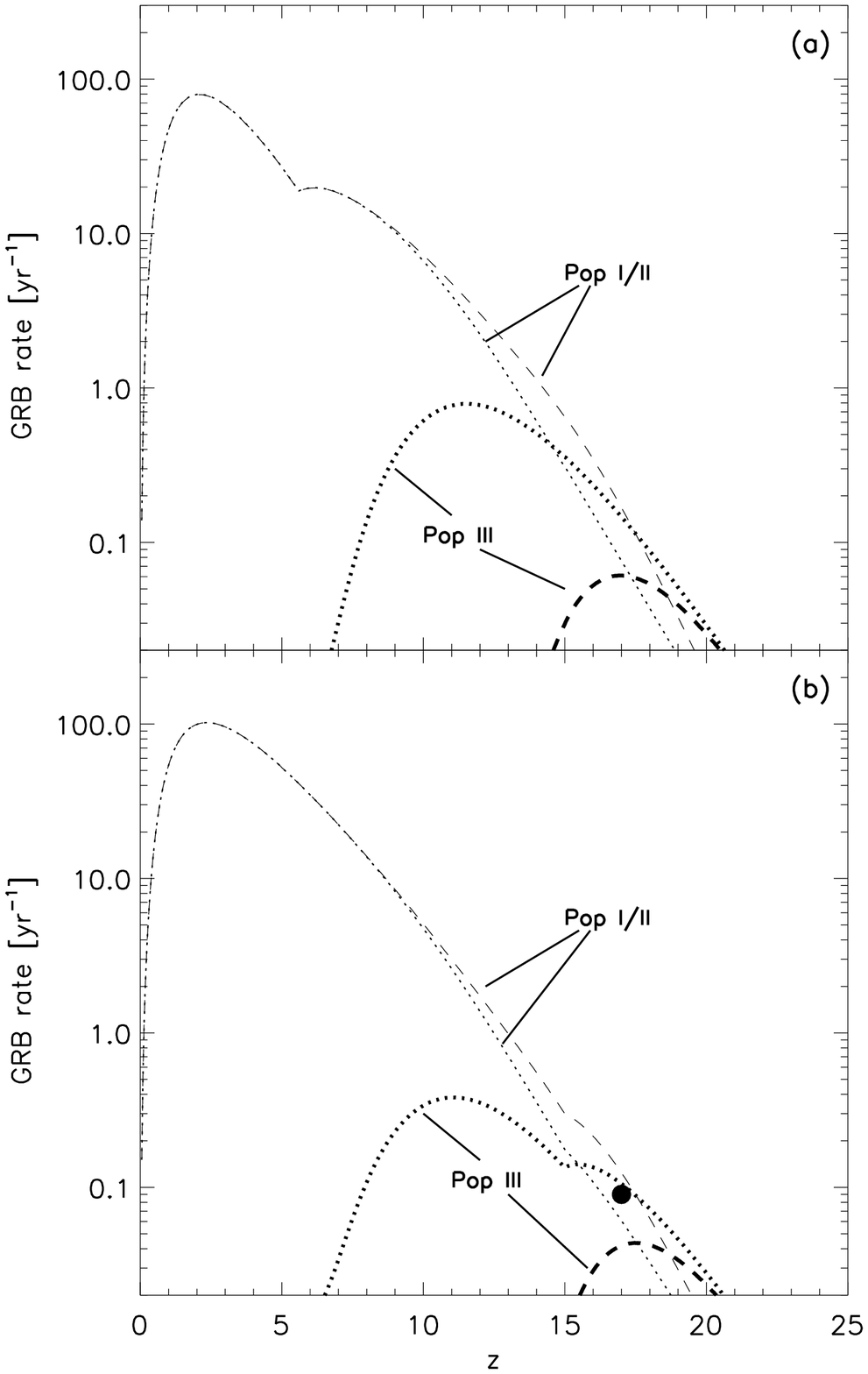}}
\caption{Predicted GRB rate observed by {\it Swift}.  Shown is the observed
number of bursts per year, $dN_{\rm GRB}^{\rm obs}/d\ln (1+z)$, as a function
of redshift.  All rates are calculated with a constant GRB efficiency,
$\eta_{\rm GRB}\simeq 2\times 10^{-9}$~bursts $M_{\odot}^{-1}$, using the
cosmic SFRs from Fig.~1.  {\it (a)} Late reionization ($z_{\rm
reion}\approx 7$).  {\it Dotted lines:} Contribution to the observed GRB
rate from Pop~I/II and Pop~III for the case of weak chemical feedback.
{\it Dashed lines:} Contribution to the GRB rate from Pop~I/II and Pop~III
for the case of strong chemical feedback.  {\it (b)} Early reionization
($z_{\rm reion}\approx 17$).  The lines have the same meaning as in panel
{\it (a)}.  {\it Filled circle:} GRB rate from Pop~III stars if these were
responsible for reionizing the universe at $z\sim 17$ (see text).  The GRB
rates from Pop~III progenitors are very uncertain; they could be zero, or,
in the other extreme, display an enhancement by up to one order of
magnitude above the baseline rates shown here (see text for details).
\label{fig2}
}
\end{inlinefigure}

Over a particular time interval $\Delta t_{\rm obs}$ in the observer
frame, the observed number of GRBs originating between redshifts $z$
and $z+dz$ is
\begin{equation}
\frac{dN_{\rm GRB}^{\rm obs}}{dz}=\psi_{\rm GRB}^{\rm obs}(z)
{\Delta t_{\rm obs}\over (1+z)}
\frac{{\rm d}V}{{\rm d}z} \mbox{\ \ ,}
\end{equation}
where ${\rm d}V/{\rm d}z$ is the comoving volume element per unit redshift
(see Bromm \& Loeb 2002). As a final step, we normalize the GRB formation
efficiency per unit mass in Pop~I/II stars to $\eta_{\rm GRB}\simeq 2\times
10^{-9}$ GRBs~$M_{\odot}^{-1}$. This choice results in a predicted number
of $\sim 90$ GRBs per year detectable by {\it Swift}. In Figure~2, we show
the {\it Swift} GRB rate, associated with Pop~I/II star formation.  For
both early and late reionization, the observed distribution is expected to
peak around $z\sim 2$. This distribution is broadly consistent with the
first GRB redshifts, still limited in number, measured during the first
months of the {\it Swift} mission (Berger et al. 2005). We now turn to the
possible contribution to the high redshift GRB rate from Pop~III stars.

\subsection{Population III Contribution}

We begin by assuming that Pop~III star formation gives rise to GRBs with
the same (constant) efficiency as is empirically derived for Pop~I/II
stars. As is evident from Figure~2, only for the case of weak chemical
feedback is {\it Swift} expected to detect a few bursts deriving from
Pop~III progenitors over the $\sim 5$~yr lifetime of the mission.  Whether
reionization occurred early or late, on the other hand, has only a small
effect on the predicted rates.  It is quite possible, on the other hand,
that {\it Swift} will not detect any Pop~III GRBs at all. Regardless of the
uncertain contribution from Pop~III stars, however, the prediction that
$\sim 10$\% of all {\it Swift} bursts originate at $z\ge 5$ is rather
robust. This fraction is due to Pop~I/II progenitors that are known to
produce GRBs, and those should exist at $z\ge 5$.

Adopting the same $\eta_{\rm GRB}$ for Pop~III as for Pop~I/II, however,
could be significantly in error. To examine the fundamental difference
between the stellar populations, we need to go beyond the phenomenological
approach pursued so far and discuss the properties of plausible GRB
progenitors in greater physical detail.

\subsubsection{Collapsar Engine}

Existing evidence implies that long-duration bursts are related to the
death of a massive star, leading to the formation of a BH (see review by
Piran 2004). 
The popular collapsar model assumes that an accretion torus is temporarily
formed around the black hole, and that the gravitational energy released
during the accretion is able to power a strong explosion
(e.g., Woosley 1993; Lee \& Ramirez-Ruiz 2005).
For the collapsar to result in a GRB, additional requirements have to be
met beyond the formation of a BH. We will discuss these next, and will then
explore whether a subset of Pop~III stars could successfully launch a GRB
under the collapsar scenario.

To successfully produce a GRB with a collapsar, three basic requirements
have to be fulfilled (see, e.g., Zhang \& Fryer 2004; Petrovic et
al. 2005):

\noindent
{\it (i)} The progenitor star has to be sufficiently massive to result in
the formation of a central BH. Collapse to a BH could occur either directly
for initial masses of the progenitor $ \ga 40 M_{\odot} $, or in a delayed
fashion by fallback of the ejecta following a failed SN explosion for
progenitor masses $25 \la M_{\ast} \la 40 M_{\odot}$ (e.g., Heger et
al. 2003). The number of BH forming stars resulting from a given total
stellar mass, here denoted by $ \eta_{\rm BH} $, will depend on the
stellar IMF which in turn is predicted to differ between the Pop~I/II and
Pop~III cases.

For simplicity, we assume that the IMF in both cases consists of a
power-law with the standard Salpeter value, $dN/dm \propto m^{-2.35}$,
but with different values for the lower and upper mass limits, $M_{\rm
low}$ and $M_{\rm up}$ respectively. For Pop~I/II stars, we take these to
be: $M_{\rm low}=0.1 M_{\odot}$ and $M_{\rm up}=100 M_{\odot}$. The
Pop~III IMF, on the other hand, is still very uncertain (see, e.g.,
Bromm \& Larson 2004). The upper mass limit can be conservatively estimated
to be $M_{\rm up}\sim 500 M_{\odot}$ (Bromm \& Loeb 2004), whereas for the
lower limit, we consider two possibilities: $M_{\rm low}\sim 30 M_{\odot}$
(e.g., Tan \& McKee 2004) and $\sim 100 M_{\odot}$ (e.g., Abel et al. 2002;
Bromm et al. 2002).

In general, 
\begin{equation}
\eta_{\rm BH}=\frac{\int_{M_{\rm BH}}^{M_{\rm up}}m^{-2.35}dm}
{\int_{M_{\rm low}}^{M_{\rm up}}m^{-1.35}dm} \mbox{\ ,}
\end{equation}
where $M_{\rm BH}\simeq 25 M_{\odot}$. For the Pop~I/II case, this results
in $\eta_{\rm BH}\simeq 1/(700 M_{\odot})$.  The Pop~III lower mass limit
exceeds the threshold for BH formation in either case, $M_{\rm low} >
M_{\rm BH}$. Not {\it every} Pop~III star, however, will leave a BH
behind. In the narrow mass range of $\sim 140 - 260 M_{\odot}$, Pop~III
stars are predicted to undergo a pair-instability supernova (PISN)
explosion (e.g., Fryer, Woosley, \& Heger 2001; Heger et al. 2003). A PISN
will lead to the complete disruption of the star, such that no compact
remnant will be left behind. For the Pop~III case, the expression results
in $\eta_{\rm BH}\simeq 1/(80 M_{\odot})$ for $M_{\rm low}=30 M_{\odot}$,
and $\eta_{\rm BH}\simeq 1/(300 M_{\odot})$ for $M_{\rm low}=100
M_{\odot}$.  Thus, the BH formation efficiency is larger for Pop~III
compared to Pop~I/II by up to one order of magnitude, depending on the
lower mass limit.

\noindent
{\it (ii)} The progenitor star has to be able to lose its hydrogen envelope
in order for the relativistic outflow to penetrate through and exit the
star (e.g., Zhang et al. 2004). This requirement derives from the observed
burst durations, $t\la 100$~s, providing an estimate for the lifetime of
the central GRB engine. The jet can therefore only travel a distance of
$r\sim c t \sim 50 R_{\odot}$ before being slowed down to non-relativistic
speeds. Massive stars with hydrogen envelopes grow to a large size during
their later evolutionary phases. For example, red supergiants can reach
radii of up to $\sim 10^3 R_{\odot}$ (e.g., Kippenhahn \& Weigert 1990).
The effectiveness of mass loss crucially depends on metallicity (e.g.,
Kudritzki 2002), and on whether the star is isolated or part of a binary
system. Below, we will discuss both effects further.

\noindent
{\it (iii)} The progenitor star has to contain a central core with
sufficient angular momentum to allow an accretion disk to form around the
growing BH. Important aspects of stellar structure and evolution can be
understood by dividing the star into a compact core, and an extended outer
envelope (e.g., Kippenhahn \& Weigert 1990). Depending on the evolutionary
stage, a radiative core is surrounded by a convective envelope, or vice
versa.  The pre-collapse stellar core has a mass $M_c$, radius $R_c$,
angular velocity $\omega_c$, and is characterized by a specific angular
momentum $j_c\simeq R_c^2 \omega_c$. Assuming that the collapse to a BH
conserves specific angular momentum, the condition for an accretion torus
to form around a growing BH in the center with mass $M_{\rm BH}$ is
centrifugal support for material orbiting at the last-stable radius.  This
condition can be expressed as $j_c \ga j_{\rm min}$, with $j_{\rm
min}=\sqrt{6}G M_{\rm BH}/c\simeq 3\times 10^{16} {\rm cm}^2 {\rm s}^{-1}$
(e.g., Podsiadlowski et al. 2004).

Recent results obtained with sophisticated stellar evolution codes that
include the effects of magnetic torques (e.g., Spruit 2002) have
demonstrated the difficulty to identify progenitor systems for
collapsar--driven GRBs that fulfill both conditions {\it (ii)} and {\it
(iii)}.  The basic problem is that the removal of the extended H-envelope
is accompanied by the loss of angular momentum in the core (e.g., Petrovic
et al. 2005). We will explore this problem next, first for single-star
progenitors, and then for binaries.

\subsubsection{Single-star Progenitor}

In massive Pop~I/II stars, radiation driven winds can lead to vigorous mass
loss, where the main source of opacity is provided by metal lines (e.g.,
Kudritzki \& Puls 2000).  Empirically, the existence of Wolf-Rayet (WR)
stars proves that Pop~I/II stars can indeed experience catastrophic mass
loss, leading to the removal of the entire hydrogen-, and, in extreme
cases, even of the helium-envelope.  The violent mass loss, however, is
accompanied by the effective removal of angular momentum from the remaining
pre-collapse core, rendering the creation of a collapsar impossible.

For massive Pop~III stars, radiatively driven winds are predicted to be
unimportant (e.g., Kudritzki 2002; Krti\v{c}ka \& Kub\'{a}t 2005).
Alternatively, mass loss could occur as
a result of stellar pulsations (through the $\epsilon$ mechanism).
Simplified, linear calculations, however, indicate that this mechanism is
not important below $\sim 500 M_{\odot}$ (Baraffe, Heger, \& Woosley 2001).
There still remains an unexplored possibility that Pop~III stars could
experience significant mass loss driven by radiation pressure on He$^{+}$
ions, where the opacity is provided by bound-free
transitions. Alternatively, processed material from preceeding episodes of
nuclear burning could be transported to the stellar surface by convection,
rotationally-induced mixing, and diffusion, thus enriching the atmosphere
to $Z\ga 10^{-4} Z_{\odot}$ at which point line--driven mass loss is
predicted to set in 
(Kudritzki 2002; Marigo, Chiosi, \& Kudritzki 2003). Recently, it has been
suggested that WR type winds, in connection with rapid rotation and the
approach to the Eddington limit, could possibly lead to significant mass loss even
for very low-metallicity stars (Vink \& de Koter 2005).
The most likely expectation, however,
is that isolated massive Pop~III stars are not able to shed much mass prior
to their final collapse.

Thus it appears likely that the majority of massive, single star
progenitors, both for Pop~I/II
and III, cannot give rise to a collapsar-driven GRB, although for different
physical reasons. We note, however, that the progenitor for the collapsar
engine is still very uncertain. In addition to the binary scenario
(see below), massive, rapidly rotating single stars have recently
been considered (e.g., Yoon \& Langer 2005; Woosley \& Heger 2005).
Our main results, based on the cosmic Pop~III SFR and the IMF-dependent
BH fraction, however, holds in this case as well.
We next turn our attention to binary-star
progenitors.

\subsubsection{Binary-star Progenitor}

Close binary systems provide a promising avenue to simultaneously meet the
requirements of strong mass loss combined with the retention of sufficient
angular momentum in the collapsing core (e.g., Lee, Brown, \& Wijers 2002;
Izzard, Ramirez-Ruiz, \& Tout 2004).
For Pop~I/II, a binary pathway to
a collapsar--driven GRB has already been suggested (e.g., Fryer et
al. 1993). The basic idea is that a close binary system, experiencing
Roche-lobe overflow (RLOF) when the primary evolves off the main sequence,
will go through a common-envelope (CE) phase, during which the hydrogen
envelope of the primary can be removed without seriously draining away the
spin of the remaining helium core (for a recent review, see Taam \&
Sandquist 2000). During the CE phase, the binary will spiral closer
together, and the corresponding loss of orbital energy will heat the
envelope with a given efficiency, often assumed to be very high:
$\alpha_{\rm CE}\simeq 1.0$ (Taam \& Sandquist 2000). The frictional energy
release during the in-spiral phase has been shown to be sufficient to unbind
the hydrogen envelope.

For a progressively tightening binary, the spin of each component is
tidally coupled to the orbital motion: $\omega_c \sim \omega_{\rm
orb}$. Since the helium core is spun up again because of the spin-orbit
coupling, it is able to retain sufficient angular momentum to fulfill the
$j_c \ga j_{\rm min}$ requirement.

\subsubsection{GRB Formation Efficiency}

In general, we can express the GRB formation efficiency for Pop~III stars
as: $\eta_{\rm GRB}\simeq \eta_{\rm BH} \eta_{\rm bin} \eta_{\rm close}
\eta_{\rm beaming} $, where $\eta_{\rm bin}$ is the binary
fraction, $\eta_{\rm close}$ the fraction of sufficiently close binaries to
undergo RLOF, $\eta_{\rm beaming}\simeq 1/50$--$1/500$ the beaming factor,
where we conservatively assume that Pop~III bursts are collimated by the BH
central engine to the same angle as Pop~I/II progenitors (the inferred
collimation angles by Frail et al. 2001; Panaitescu \& Kumar 2001, are
indeed comparable to those of radio jets from the much more massive BHs in
galactic nuclei).
We have already discussed $\eta_{\rm BH}$, and how Pop~III star formation
is characterized by an enhancement of up to one order of magnitude because
of the higher fraction of BH--forming progenitors.

It is currently not known whether Pop~III stars can form binaries, and if
so, what the corresponding binary fraction will be (e.g.,
Saigo, Matsumoto, \& Umemura 2004).
In present-day star
formation, the incidence of binaries is high, with $\sim 50$\% of all stars
occurring in binaries or small multiple systems (e.g., Duquennoy \& Mayor
1991). 
Current three-dimensional simulations of the formation of the first
stars still lack the resolution to resolve the possible fragmentation
of a collapsing cloud into {\it tight} binaries or multiple stellar cores
on scales $\la 100$~AU.
Although we cannot yet conclusively
address the formation of close binaries, there is evidence from numerical
simulations that binary or multiple clump formation is rather common on
larger scales ($\ga 0.1$~pc).
In simulations where the collapsing
gas had acquired a high degree of angular momentum, and where the collapse
led to a disk-like configuration, pre-stellar clumps commonly occurred
in binary or multiple systems (e.g., Bromm et al. 1999, 2002; Bromm \& Loeb
2003a, 2004).
Thus motivated, 
our best guess is that $\eta_{\rm bin}\la 0.5$. More work, in particular
involving improved numerical simulations, is required to constrain this
crucial quantity further.

Provided that Pop~III star formation does include a fraction of binaries,
we can use the collapsar requirements to obtain an estimate for the maximum
binary separation, $a_{i, {\rm max}}$, prior to the CE inspiral phase, as
follows. Assuming for simplicity that a Pop~III binary has equal mass
components, the Roche radius is $r_L \sim 0.5 a_i$. A CE phase will only
occur when the star overflows its Roche lobe during the red giant phase:
$R_{\rm RG} > r_L \ga a_i$. We estimate the Pop~III radius during the giant
phase, which is smaller than that of a Pop~I star of equal mass, to be
$R_{\rm RG}\sim 300 R_{\rm MS}$, where $R_{\rm MS}$ is the main-sequence
radius. Massive Pop~III stars obey a simple mass-radius
relation (Bromm et al. 2001b): $M\propto R_{\rm MS}^2$. The maximum
separation for RLOF and therefore a CE phase to occur is thus
\begin{equation}
a_{i, {\rm max}} \simeq 10^3 R_{\odot}\left(\frac{M}{10^2 M_{\odot}}
\right)^{1/2} \mbox{\ .}
\end{equation}

The minimum possible binary separation, $a_{i, {\rm min}}$, on the other hand,
will determine the extent of the inspiral process and therefore of the
accompanying frictional heating of the hydrogen envelope as well as the
tidal spinning up of the helium core. We approximately assume that the
minimum separation is given by twice the radius of a massive Pop~III star
during the main--sequence phase (which is only weakly dependent on mass):
$a_{i, {\rm min}} \simeq 10 R_{\odot}$. If we further assume that the
initial separations are distributed with equal probability per logarithmic
separation interval, as they are for Pop I/II binaries (e.g., Abt 1983;
Heacox 1998; Larson 2003), $dN/d\ln a \propto {\rm const.}$, and that the
largest possible binary separation is given by the Jeans length for the
typical conditions in a primordial gas cloud, $\lambda_{J}\simeq 1$~pc
(e.g., Bromm et al. 2002, Bromm \& Loeb 2004), we estimate the fraction of
all Pop~III binaries that are sufficiently close to experience a CE phase
to be $\eta_{\rm close} \sim 30$\%.

In summary, $\eta_{\rm GRB}$ for Pop~III stars is very uncertain, and could
be zero in case that no Pop~III binaries existed. On the other hand, one
may argue that the binary properties for Pop~III had been similar to
Pop~I/II, in cases where Pop~III star formation takes place in more massive
host systems that could give rise to a stellar cluster, or at least
multiple stars.  Such a clustered environment is often suggested to explain
the formation and the properties of binaries in present-day galaxies (e.g.,
Larson 2003).  Assuming that the Pop~III binary properties are similar to
Pop~I/II, we find a significant enhancement in $\eta_{\rm BH}$ due to the
difference in the IMF between the populations. We can then place an upper
limit on the GRB rate from Pop~III stars by multiplying the baseline rates
in Figure~2 by a factor of $\sim 10$. This would result in Pop~III GRB
rates as large as $\sim 10$~bursts detected by {\it Swift} per year
for the case of weak chemical feedback. Such
very high rates can already be excluded, since {\it Swift} has only
identified two GRBs from $z\ga 5$ as of yet, GRB~050814 at $z\simeq 5.3$
(Jakobsson et al. 2005)
and GRB~050904 at $z\simeq 6.3$
(Antonelli et al. 2005; Haislip et al. 2005;
Kawai et al. 2005). For strong chemical feedback, on the other hand,
we predict rates of less than one burst detected per year, even if
the BH efficiency were increased by one order of magnitude, and such
a Pop~III contribution cannot be excluded with the current constraints
from {\it Swift}.

\subsection{Constraints from Reionization}

If massive Pop~III stars led to an early reionization of the universe at
$z\sim 17$, as may be required by the {\it WMAP} data (Kogut et al. 2003),
we can obtain an estimate for the corresponding Pop~III SFR at $z_{\rm
reion}$ and for the possible accompanying GRB rate, as follows.

Pop~III stars with masses $\ga 100 M_{\odot}$ produce $\sim
10^{62}$~H--ionizing photons per solar mass over their $\sim 2\times
10^{6}$~yr lifetime (e.g., Bromm et al. 2001b).  We can then estimate that
$\sim 3\times 10^{5} M_{\odot}$ in Pop~III stars are required to produce
$\sim 5$ ionizing photons for every hydrogen atom in a comoving Mpc$^{3}$.
This number is sufficient to compensate for recombinations at the mean
cosmic density.
Assuming further that the burst of Pop~III star formation is spread over a
fraction $\epsilon$ of the Hubble time at $z_{\rm reion}\sim 17$, $\Delta
t_{\rm SF}\sim 4\times 10^{7}(\epsilon/0.2)$~yr, the comoving Pop~III SFR
able to reionize the universe at that redshift is: ${\rm SFR}_{\rm
reion}\sim 7.5\times 10^{-3}/(\epsilon/0.2)$
$M_{\odot}$~yr$^{-1}$~Mpc$^{-3}$. The extremely rapid growth of the
collapsed fraction of baryons with redshift implies a value of $\epsilon\ll
1$; however, the minimum value of $\epsilon$ is $\sim 0.1$ because even
within a single dark matter halo, star formation cannot be synchronized to
better than the dynamical time which amounts to $\epsilon \sim 0.1$ for a
virial density contrast of $\sim 200$.

In Figure~2, we show the GRB rate which would correspond to ${\rm SFR}_{\rm
reion}$ for $\epsilon =0.2$ when the constant Pop~I/II GRB efficiency
factor is used, resulting in $\sim 0.1$~GRBs per year.
Under this conservative assumption, {\it Swift} is not expected to detect,
within its expected 5 year mission lifetime,
any bursts connected to the Pop~III stars that were responsible for an
early reionization of the universe. Again, the prospects for detection
would be significantly improved if the Pop~III GRB efficiency is boosted
due to the increased fraction of BH--forming progenitors.

\section{Summary and Conclusions}

Figure 2 leads to the robust expectation that $\sim 10$\% of all {\it
Swift} bursts should originate at $z\ga 5$. This prediction is based on the
likely existence of Pop~I/II stars in galaxies that were already
metal--enriched at these high redshifts.
Additional GRBs could be triggered
by Pop~III stars, with a highly uncertain efficiency. Assuming that
long-duration GRBs are produced by the collapsar mechanism, a Pop~III star
with a close binary companion provides a plausible GRB progenitor. We
have estimated the Pop~III GRB efficiency, reflecting the probability of
forming sufficiently close and massive binary systems, to lie between zero
(if tight Pop~III binaries do not exist) and $\sim 10$ times the
empirically inferred value for Pop~I/II (due to the increased fraction of
BH--forming progenitors among Pop~III stars).

Recently, 
Gorosabel et al. (2004) and Natarajan et al. (2005) predicted the expected
redshift distribution of long-duration GRBs, assuming they trace the cosmic
star formation history, with various phenomenological prescriptions for
the dependence on metallicity. In difference from these studies, we
isolate the zero-metallicity (Pop~III) stars and treat them as potential
GRB progenitors based on a physical model.

It is of great importance to constrain the Pop~III star formation mode, and
in particular to determine down to which redshift it continues to be
prominent. The extent of the Pop~III star formation will affect models of
the initial stages of reionization (e.g., Wyithe \& Loeb 2003a,b; Ciardi,
Ferrara, \& White 2003; Sokasian et al. 2004; Yoshida et al. 2004; Alvarez,
Bromm, \& Shapiro 2006) and metal enrichment (e.g., Mackey et al. 2003;
Furlanetto \& Loeb 2003, 2005; Scannapieco et al. 2003; Schaye et al. 2003;
Simcoe, Sargent, \& Rauch 2004), and will determine whether planned surveys
will be able to effectively probe Pop~III stars (e.g., Scannapieco et
al. 2005).  The constraints on Pop~III star formation will also determine
whether the first stars could have contributed a significant fraction to
the cosmic near-IR background (e.g., Santos, Bromm, \& Kamionkowski 2002;
Salvaterra \& Ferrara 2003; Dwek, Arendt, \& Krennrich 2005;
Kashlinsky 2005; Madau \& Silk 2005;
Fernandez \& Komatsu 2006).
If Pop~III binaries were common, {\it Swift} might be the first instrument
to detect Pop~III stars from galaxies at redshifts $z\ga 7$.

\acknowledgments 

We thank Vicky Kalogera and Steve Furlanetto for helpful discussions.  VB
thanks the Institute for Theory and Computation at the CfA where part of
this work was carried out.  This work was supported in part by NASA {\it
Swift} grant NNG05GH54G.

\end{document}